\documentstyle[psfig,amssymb]{mn}
\title[The XMM-Newton/2dF survey - V. the radio properties of the X-ray population]
{The XMM-Newton/2dF survey - V. the radio properties of the X-ray population}
 
\author[A. Georgakakis et al.] {A. Georgakakis$^{1}$\thanks{email:
  age@astro.noa.gr}, I. Georgantopoulos$^{1}$,  I. Leonidaki$^{1,2}$,
  Akylas$^{1,3}$,  G. C. Stewart$^4$, 
  \\ \\
  {\LARGE C. Goudis$^{1,2}$} \\ \\  
  $^1$ Institute of Astronomy \& Astrophysics, National Observatory of
  Athens, I. Metaxa \& V. Pavlou, Athens, 15236, Greece \\ 
  $^2$ Astronomical Laboratory, Department of Physics, University of
  Patras, 26500, Patras, Greece\\
  $^3$ Physics Department University of Athens, Panepistimiopolis,
  Zografos, 15783, Athens, Greece \\	  
  $^4$ Department of Physics and Astronomy, University of Leicester
  Leicester LE1 7RH, UK \\
}
\begin{document}
\maketitle  

\begin{abstract}
In this paper we cross-correlate the FIRST 1.4\,GHz radio survey
with a wide field ($\rm 1.6\,deg^{2}$) shallow [$f_X\rm (0.5 - 
8\,keV)\approx 10^{-14} \, erg \, s^{-1}\, cm^{-2}$] XMM-{\it Newton}
survey. We find 12 X-ray/radio matches representing 4\% of the X-ray
selected sample. Most of them are found to be associated with AGNs
(total of 9) on the basis of the observed optical spectra (3), radio
morphology (2) or X-ray/optical properties  (4), while  one radio 
source is identified with an X-ray selected cluster. We also find two
sources associated with low redshift galaxies with narrow emission
line  optical spectra, X-ray luminosity $L_X(\rm 0.5-8\,keV)\rm
\approx 10^{41}\,erg\,s^{-1}$,  radio luminosity density $L_{\rm
1.4\,GHz}\rm \approx  5\times10^{22}\,W\,Hz^{-1}$ and $\log
f_X/f_{opt} \approx -2$  
suggesting `normal' star-forming galaxies. We argue that radio surveys 
combined with X-ray samples could provide a powerful tool for
identifying X-ray selected `normal' galaxies powered by stellar
processes. Finally, radio loud and quiet systems in the present sample 
have mean X-ray spectral properties consistent with
$\Gamma\approx1.9$.  
\end{abstract}

\begin{keywords}  
  Surveys -- Galaxies: active -- X-rays: galaxies -- X-rays: general 
\end{keywords} 

\section{Introduction}\label{sec_intro}
X-ray surveys are not only dominated by AGNs but are also believed to
be the least biased method for selecting such systems (e.g. Brinkmann
et al. 2000). The evidence above has motivated a number of studies
aiming to compile X-ray selected (and thus least biased) AGN samples
to understand the large variations in their observational properties
and to test  unification schemes.   At radio wavelengths in particular
large programs have been initiated by cross-correlating  wide area
radio surveys with archival X-ray data aiming to understand the radio
loud/quiet AGN dichotomy (Brinkmann et al. 1997; Caccianiga et
al. 1999; Brinkmann et al. 2000).  

For example Brinkmann et al. (2000) cross-correlated the FIRST radio
survey (Faint Images of the Radio Sky at Twenty centimeters; Becker et 
al. 1995) with the ROSAT All Sky Survey (RASS-II; Voges et al. 1999) at a
0.1-2.4\,keV flux limit of $\approx10^{-13}\rm\, erg\,
s^{-1}\,cm^{-2}$. About 1/3 of the X-ray sources are identified with
radio counterparts.  The above fraction of X-ray/radio associations is 
higher than that found in previous soft X-ray selected samples
(Stocke et al. 1991; Zamorani et al. 1999; Ciliegi et al. 1995; see
Ciliegi et al. 2003 for a summary), most likely due to the 
relative X-ray and radio limits of the  RASS-II and the FIRST
surveys. Follow-up  optical spectroscopy of the X-ray/radio matches
shows that the majority of these sources are AGNs spanning both the
radio loud and the radio quiet regime. These data indicate that there
is no sharp boundary between radio loud and radio quiet AGNs but
rather a smooth transition. A highly significant X-ray--radio
luminosity correlation for these sources is also reported. Apart from
AGNs the sample of Brinkmann et al. (2000) also comprises a small
fraction ($\approx5\%$) of `normal' galaxies likely to be starbursts.  

At fainter X-ray flux limits ($\approx10^{-16}\rm\, erg\, s^{-1}\,
cm^{-2}$) Bauer et al. (2002) investigated the radio  properties of
the X-ray sources using data from the 1\,Ms Chandra Deep Field North
(CDF-N) combined with  $\mu$Jy sensitivity VLA  observations. They
find that $\approx38\pm20\%$ of the X-ray selected AGNs have radio
counterparts  to the limit of the VLA data (0.05\,mJy). They also
argue that the radio emission, at least in the sub-sample of harder
(i.e. obscured) AGNs, is likely to be associated with circumnuclear
star-formation activity.  Contrary to AGNs the largest overlap between
X-ray and radio sources ($66\pm25\%$) is for the sub-sample of narrow
emission line galaxies likely to be star-forming  systems. The
evidence above suggests the emergence of `normal' and starburst
galaxies at faint X-ray fluxes.    

In addition to soft X-ray selected samples a number of hard X-ray
surveys have been followed by radio observations yielding X-ray/radio
matched samples with identification rates  $\ga 30\%$
(e.g. Ciliegi et al. 2003; Akiyama et al. 2000; Barger et
al. 2001; Georgakakis et al. 2004b). Ciliegi et al. (2003) argue that
this is likely due to observational effects (e.g. deeper radio data) and
the fact that both the radio wavelengths and the hard X-ray energies
($\rm >2\,keV$) are least biased to obscuration effects.

The studies above mainly concentrate on either very bright
($>10^{-13}\rm \, erg \, s^{-1} \, cm^{-2}$) or extremely faint
($<10^{-15}\rm \, erg \, s^{-1} \, cm^{-2}$) X-ray sources. In the
present study we explore the radio  properties of the X-ray population
in the intermediate X-ray flux range ($\approx10^{-14}\rm\, erg\,
s^{-1}\, cm^{-2}$). We cross correlate the FIRST radio  survey with a
wide area ($\rm 1.6\,deg^{2}$) shallow [$f_X(\rm
0.5-8\,keV)\approx10^{-14}\rm \, erg\, s^{-1}\, cm^{-2}$] XMM-{\it
Newton} survey near the North  Galactic Pole region (XMM-{\it
Newton}/2dF survey\footnote{http://www.astro.noa.gr/$\sim$xray/}).     

Section \ref{sec_obs} describes the X-ray, radio and optical data used
in the present study while section \ref{sample} presents the X-ray/radio
matched sample. The results are discussed in section \ref{discussion}
with section \ref{conclusions} summarising our conclusions.
Throughout this paper 
we adopt $\rm H_{o}=65\,km\,s^{-1}\,Mpc^{-1}$, $\rm \Omega_{M}=0.3$
and  $\rm \Omega_{\Lambda}=0.7$.   

\section{Observations}\label{sec_obs}
\subsection{X-ray data}\label{sec_obs_xray}
The X-ray sample used in the present study is compiled from the
XMM-{\it Newton}/2dF survey. This is a wide area ($\rm \approx
2.5\,deg^2$) shallow [$f_X(\rm 0.5 - 8 \,keV) \approx 10^{-14} \,erg
\,s^{-1} \,cm^{-2}$; $5\sigma$] survey carried out by the XMM-{\it
Newton} near the North and the South  Galactic Pole regions. The data
reduction, source  extraction, flux estimation  and catalogue
generation are described in detail by Georgakakis et al. (2003a, Paper
\,I; 2004a, Paper\,II). In the present study we 
concentrate on the North Galactic Pole F864  subregion of the
XMM-{\it Newton}/2dF survey covering an area of $\rm 1.6\,deg^2$. This
is because of the wealth of follow-up observations (optical photometry
and spectroscopy, radio) available for these fields. The X-ray sample
comprises a total of 291 sources detected in the 0.5-8\,keV spectral
band above the $5\sigma$ threshold. We note that about 10\% of the
surveyed area is covered at the flux $f_X(\rm 0.5 - 8 \,keV) =
10^{-14} \,erg \,s^{-1} \,cm^{-2}$. This fraction increases to about
50\% at $f_X(\rm 0.5 - 8 \,keV) = 2\times10^{-14} \,erg \,s^{-1}
\,cm^{-2}$.  

\subsection{Radio data}\label{sec_obs_radio}
The radio data are from the FIRST (Faint Images of the Sky at Twenty
centimeters) radio survey (Becker et al. 1995; White et
al. 1997). The observations are carried out at 1.4\,GHz using the NRAO
Very Large Array (VLA) in the B-configuration.  The limiting flux
density  is $S_{\rm 1.4GHz}=0.8\rm\,mJy$ with a $5\sigma$ source
detection limit of $\rm \approx1\,mJy$. The
catalogue is estimated to be 95 and 80\% complete at 2 and 1\,mJy
respectively (Becker et al. 1995). The 
positional accuracy of the detected sources is better than
$\approx1$\,arcsec facilitating their optical identification.   

The radio data used in the present study are from the April 11, 2003
version of the FIRST VLA catalogue. A total of 78 radio sources
overlap with the XMM-{\it Newton} pointings of the F864 region.

\subsection{Optical photometric and spectroscopic data}\label{sec_obs_opt}
The XMM-{\it Newton}/2dF survey F864 region overlaps with
the Sloan Digital Sky Survey (York et al. 2000). The  SDSS is an
on-going imaging and spectroscopic survey that aims to cover about $\rm
10\,000\,deg^2$ of the sky. Photometry is performed in 5 bands
($ugriz$;  Fukugita et al. 1996; Stoughton et al. 2002) to the
limiting magnitude $g \approx 23$\,mag, providing a uniform and
homogeneous multi-color photometric catalogue. The SDSS spectroscopic
observations will obtain spectra for over 1 million objects, including
galaxies brighter than $r=17.7$\,mag, luminous red galaxies to
$z\approx0.45$ and colour selected QSOs (York et al. 2000; Stoughton
et al. 2002). In the present study we use data from the Early Data
Release (EDR; Stoughton et al. 2002). 

In addition to the SDSS the F864 region overlaps with the
recently completed 2dF Galaxy Redshift Survey
(2dFGRS\footnote{http://msowww.anu.edu.au/2dFGRS/}; Colless et 
al. 2001; Colless et al. 2003) and the 2dF QSO Redshift Survey
(2QZ\footnote{http://www.2dfquasar.org}; Croom et al. 2001). Both the
2dFGRS and 2QZ are large-scale spectroscopic campaigns that fully
exploit the capabilities of the 2dF multi-fibre spectrograph on the
4\,m Anglo-Australian Telescope (AAT). These
projects provide high quality spectra, redshifts and spectral
classifications for 220\,000 $bj<19.4$\,mag galaxies and 23\,000
optically selected $bj<20.85$\,mag QSOs.

\begin{table*} 
\footnotesize 
\begin{center} 
\begin{tabular}{llcc cccc ccc}
\hline 
ID &
NAME &
$\alpha_{\rm 1.4GHz}$ & 
$\delta_{\rm 1.4GHz}$ & 
$\delta\theta(\rm RX)$ &
$\delta\theta(\rm RO)$ &
$P_{RO}$ &
$g$    &
$S_{\rm 1.4GHz}$    &
$z$    &
Class$^a$  
\\
 
 &
 &
(J2000)&
(J2000)&
(arcsec) &
(arcsec) &
(\%)     &
(mag)    &
(mJy)    &
         &
         \\

1$^{b}$ &
 FIRST\,J134304.6--000055 &
 13 43 04.61 & --00 00 55.35 &
 3.3  &   0.9 & 0.27 &
 21.97 &
 28.26 &
  -- &
  DOUBLE 
 \\

2 &
FIRST\,J134233.0--001553 &
13 42 33.04 & --00 15 53.24 &
 3.6   &  2.9 & 0.44 &
 19.98 &
 0.73 &
 $2.132^{3}$ &
 BL\\

3 &
 FIRST\,J134212.2--001737 &
 13 42 12.21 &  --00 17 37.82 &
 4.3  &   0.1 & $<0.01$ &
 17.18 &
 1.50 &
 $0.087^2$ &
 EA\\

4 &
 FIRST\,J134347.5+002024 &
 13 43 47.47 & +00 20 24.22 &
  2.9  &   0.5 & 0.01 &
 19.55 &
 3.13  &
 $0.240^1$&
 AB\\

5 &
 FIRST\,J134414.2+001642 &
 13 44 14.18 & +00 16 42.31 &
  1.1  & 0.4 & 0.01 & 
  19.67 &
  95.96 &
  -- &
  DOUBLE 
  \\

6 &
 FIRST\,J134232.4--003151 &
 13 42 32.38 & --00 31 51.10 &
 0.4  &   0.1 &  $<0.01$ &
 18.98 &
 1.55 &
 $1.209^{4}$  &
 BL \\

7 &
 FIRST\,J134133.4--002432 &
 13 41 33.36 & --00 24 32.26 &
 2.8  &   0.3 &  $<0.01$ &
 17.08 &
 2.98 &
 $0.072^2$ &
 NL \\

8 &
 FIRST\,J134137.7--002555 &
 13 41 37.68 & --00 25 55.02 &
 2.7  & 0.3 & $<0.01$ &
 16.56 &
 2.37 &
 $0.052^{1,5}$ &
 NL \\

9 &
 FIRST\,J134412.9--003006 &
 13 44 12.89 & --00 30 05.51 &
 2.5  &   0.1 & $<0.01$ &
 20.55 &
 8.03  &
 $0.708^{4}$ &
 BL \\

10 &
 FIRST\,J134431.8--002832 &
 13 44 31.77 & --00 28 31.48 &
 2.9  &  0.6 & 0.50 &
 23.80 &
 1.43 &
 -- &
 -- \\

11 &
 FIRST\,J134128.4--003120 &
 13 41 28.36 & --00 31 20.21 &
 0.2 & 0.4 & 0.04 &
 21.55 &
 0.81 &
 -- & 
 -- \\

12 &
 FIRST\,J134447.0--003009 &
 13 44 46.97 & --00 30 09.29 &
 0.7  &  0.7 & 0.22 &
 22.30  & 
 50.46  &
  -- &
  DOUBLE 
 \\
 
\hline
\multicolumn{11}{l}{$^a$AB: absorption lines; NL: Narrow emission lines;
BL: Broad optical emission lines; EA: absorption+emission lines}\\
\multicolumn{11}{l}{DOUBLE: double lobe morphology} \\
\multicolumn{11}{l}{$^b$this double lobe radio sources is associated
with an X-ray cluster.}\\ 
\multicolumn{11}{l}{1: XMM-{\it Newton}/2dF spectroscopic program; 2:
2dFGRS; 3: 2QZ; 4: SDSS; 5: Terlevich et al. (1991)}\\ 
\end{tabular} 
\end{center} 
\caption{
X-ray/radio matched sources in the XMM-{\it Newton}/2dF survey.  
}\label{tbl2} 
\normalsize  
\end{table*}

\begin{table*} 
\footnotesize 
\begin{center} 
\begin{tabular}{lc ccc cc} 
\hline 
ID &
net counts &
$f_X(\rm 0.5-8.0\,keV)$ &
$L_X(\rm 0.5-8.0\,keV)$ &
$L_{\rm 1.4 GHz}$ &
$\rm N_H$ &
$\Gamma$
 \\

  &
(0.5-8.0\,keV) &
($\times 10^{-14}\,\rm erg\,s^{-1}\,cm^{-2}$) &
($\rm\,erg\,s^{-1}$) &
($\rm\,W\,Hz^{-1}$) &
($\rm 10^{20}\,cm^{-2}$) &
    
\\
    
%

1 & 
45 &
 $4.36\pm0.98$ &
 -- &
 -- &
 $<7.5$ &
 $1.81^{+0.35}_{-0.32}$
\\

2 &
24 & 
 $2.78\pm1.03$&
 $(8.74\pm3.24)\times10^{44}$ &
 $7.19\times10^{25}$ &
 $<30$ &
 $1.48^{+0.50}_{-0.54}$
\\

3 &
21 &
  $3.64\pm1.08$ &
  $(7.89\pm2.34)\times10^{41}$ &
  $3.53\times10^{22}$  &
   $<3.0$ &
  $2.82^{+1.51}_{-1.05}$
\\

4$^a$ &
141 &
 $26.9\pm2.34$ &
 $(5.21\pm0.45)\times10^{43}$ &
 $7.52\times10^{23}$ &
 $<0.6$ &
 $2.03^{+0.16}_{-0.23}$
\\

5$^{b}$ &
766 &
  $93.8\pm3.82$ &
  -- &
  -- &
 $<1.4$ &
 $1.76^{+0.19}_{-0.13}$
\\

6 &
93 &
 $11.10\pm1.45$ &
 $(9.25\pm1.21)\times10^{44}$ & 
 $2.85\times10^{25}$ &
 $<1.4$ &
 $2.60^{+0.41}_{-0.37}$
\\

7 &
22 &
  $1.31\pm0.43$ &
  $(1.90\pm0.63)\times10^{41}$ & 
  $4.64\times10^{22}$ &
  $<4$ &
  $2.47^{+0.99}_{-0.78}$
\\

8 &
30 &
  $1.95\pm0.52$ &
  $(1.44\pm0.39)\times10^{41}$ &
  $1.84\times10^{22}$ &
  $<8.7$ &
  $2.14^{+0.70}_{-0.46}$
\\

9 &
77 &
 $6.11\pm0.8$ &
 $(1.44\pm0.19)\times10^{44}$ &
 $3.23\times10^{25}$ &
 $<0.3$ &
 $2.80^{+0.22}_{-0.26}$
\\

10 &
31 &
 $1.67\pm0.40$ &
 -- &
 -- &
 $<4$ &
 $2.27^{+0.48}_{-0.42}$
\\

11 &
110 &
 $9.09\pm0.96$ &
 --&
 --&
 $150^{+50}_{-50}$ &
 $0.51^{+0.19}_{-0.28}$
\\

12 &
32 &
 $1.95\pm0.45$ &
 --&
 --&
 $56^{+61}_{-54}$ &
 $0.88^{+0.42}_{-0.42}$
\\
\hline
\multicolumn{7}{l}{$^a$X-ray flux and X-ray spectral properties are
estimated from the MOS} \\ 
\multicolumn{7}{l}{$^b$X-ray spectral fitting used $\chi^2$ statistic
with both $\Gamma$ and $\rm N_H$ free parameters}  \\
\end{tabular} 
\end{center} 
\caption{X-ray properties of the X-ray radio matched  sources
in the XMM-{\it Newton}/2dF survey}\label{tbl3}  
\normalsize  
\end{table*} 

\section{X-ray -- radio correlation}\label{sample}
In the present study we use a total of 291 XMM-{\it Newton}/2dF survey
sources with detection significance $>5\sigma$ in the 0.5-8\,keV
band. The limiting flux of these observations in the 0.5-8\,keV band
is $f_X(\rm 0.5 - 8 \,keV) \approx 10^{-14} \,erg \,s^{-1} \,cm^{-2}$.  

The FIRST and the XMM-{\it Newton}/2dF source catalogues are cross
correlated using an initial matching radius of 10\,arcsec. This is to
include in our sample well separated double lobe radio sources that
may be missed if a smaller radius is employed. The candidate
identifications were then visually inspected rejecting single 
component sources (i.e. not double lobes) that lie more than
7\,arcsec from an X-ray detection.  The final sample comprises 
a total of 12 X-ray sources  with radio counterparts. Given the
surface density of the FIRST survey ($\rm \approx 90 \,deg^2$; Becker 
et al. 1995) and the XMM-{\it Newton}/2dF sources we estimate about
0.3 spurious identifications for the above sample within the 7\,arcsec 
radius. We note that increasing the initial matching radius to
40\,arcsec does not include any additional sources in the sample.

\begin{figure*} 
\centerline{\psfig{figure=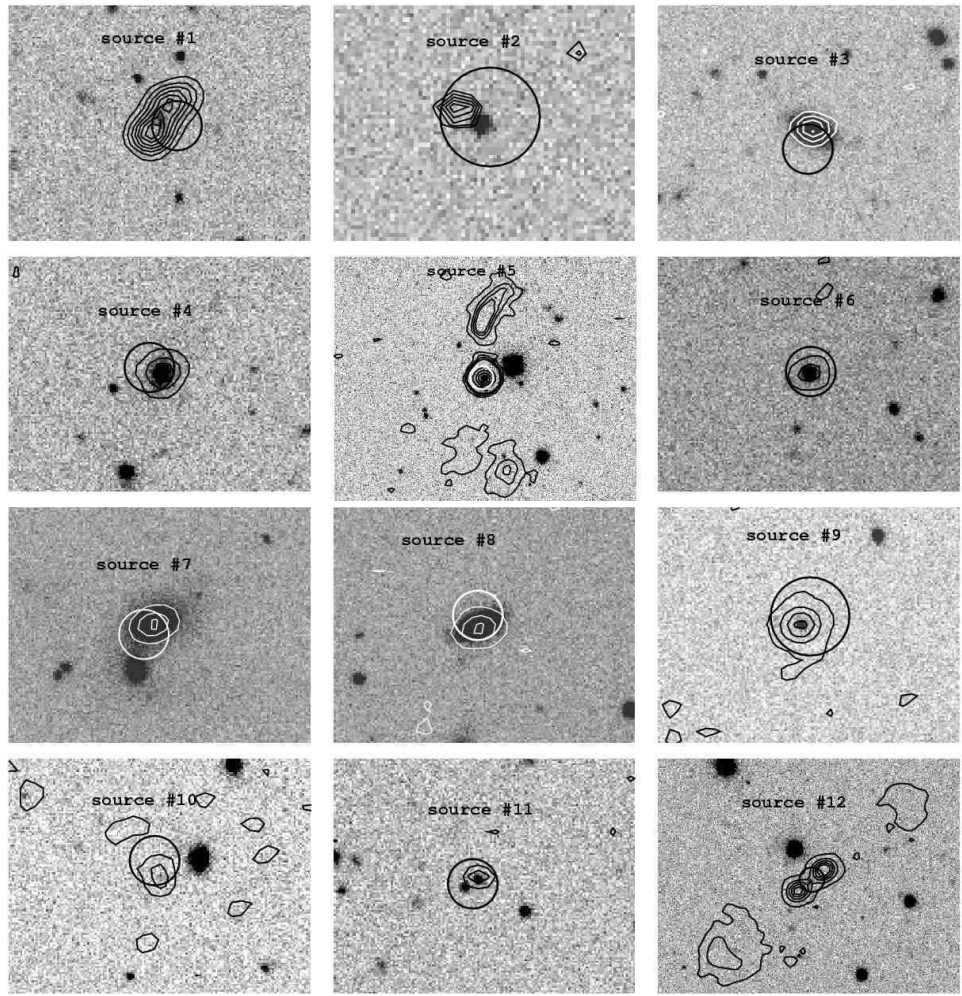,width=5in,angle=0}} 
\caption{
SDSS optical $r$-band images of the X-ray/radio matched sources in
Table \ref{tbl2} with the radio contours overlaid. The position of the
X-ray centroid is shown with the circle with a radius of
5\,arcsec. 
 }\label{fig_optical}
\end{figure*}

\begin{figure} 
\centerline{\psfig{figure=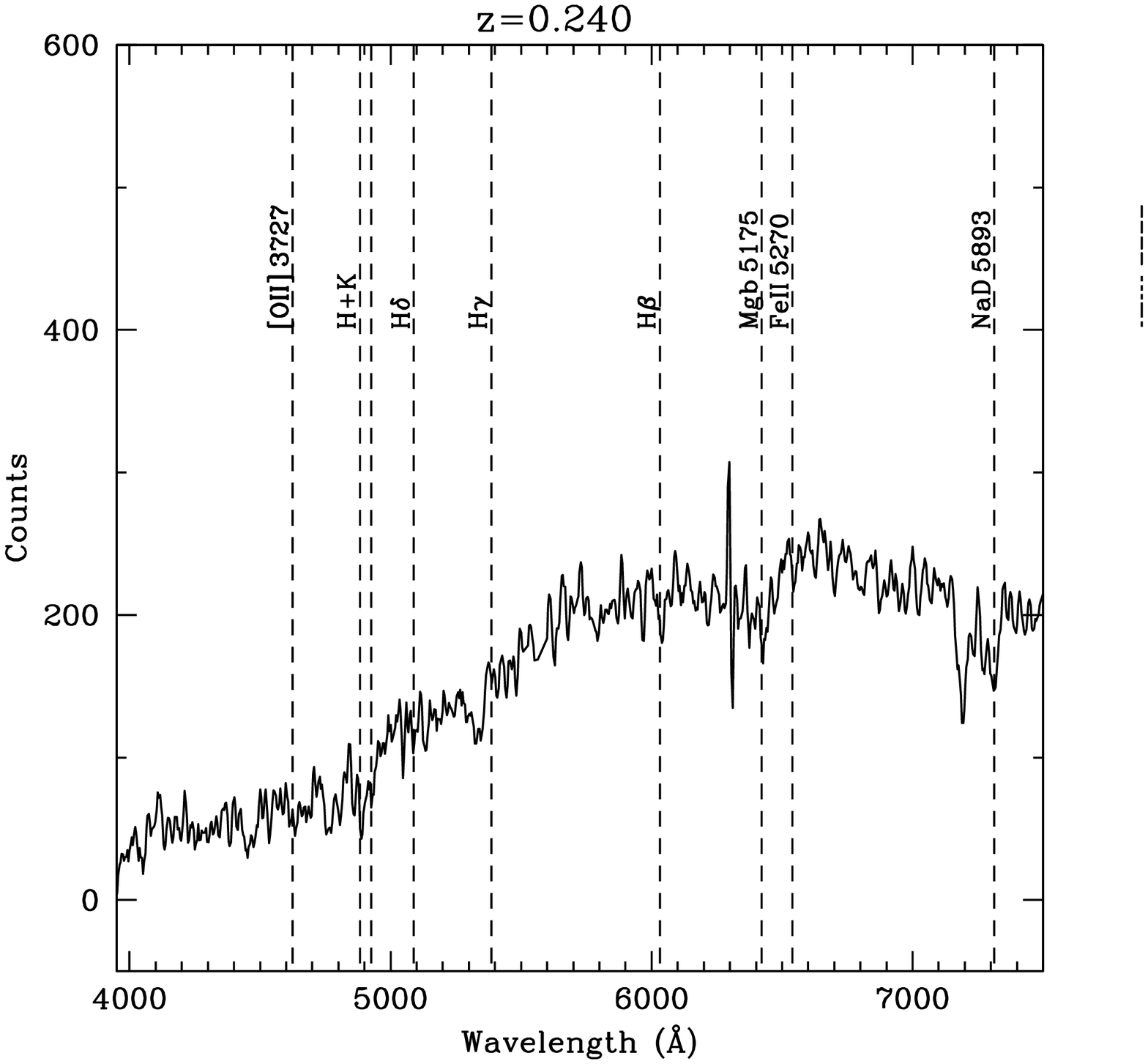,width=3.5in,angle=0}} 
\caption{
Optical spectrum of source \#4 showing absorption lines only  at
$z=0.240$.  
 }\label{fig_spec}
\end{figure}

The SDSS is used to optically identify these sources by
estimating the probability a given candidate is chance coincidence
(Downes et al. 1986).
The more  accurate radio positions are used for the optical
identifications.   Of the 12 X-ray/radio matches we propose 10
candidate optical identifications. Optical images of these sources are
shown in Figure \ref{fig_optical}. The optical spectrum of source \#4
(see Table \ref{tbl2}) is plotted in Figure \ref{fig_spec}. The
optical 
spectra of the remaining sources are either presented in Paper\,II or
can be found in the 2QZ, 2dFGRS and SDSS archives. 

The X-ray/radio matched sample is
presented in Table \ref{tbl2} which has the following format:

{\bf 1.} source identification number.

{\bf 2.} Name of the FIRST radio source closest  to the X-ray source. 

{\bf 3-4.} Right ascension ($\alpha_{\rm 1.4GHz}$) and declination
($\delta_{\rm 1.4GHz}$) of the radio source position in J2000. In the
case of double lobes we list the position of the centroid. 

{\bf 5.} Offset in arcseconds between the X-ray source centroid estimated 
by the {\sc ewavelet} task of {\sc sas} and the radio source centre. We 
note that the X-ray source centroid does not always coincide with 
the peak of the X-ray emission. 

{\bf 6.} Offset in arcseconds between the radio source position 
and the optical source centre. 

{\bf 7.} Probability, $P$, the optical identification or a radio source
is a chance coincidence.

{\bf 8.} Optical $g$-band magnitude.

{\bf 9.} 1.4\,GHz radio flux density. 

{\bf 10.} Redshift. The source from which the redshift estimate was
obtained is also listed in Table \ref{tbl2}. 

{\bf 11.} Classification on the basis of the observed optical spectral
features or radio morphology: {\bf AB:}  absorption lines only, {\bf
NL:} narrow emission lines, {\bf EA:} both absorption and emission
lines, {\bf BL:} broad emission lines and  {\bf DOUBLE:} for radio
sources with double lobes. We note that source \#1 has double lobe
morphology and is associated with an X-ray selected cluster. 

To explore the X-ray spectral properties of the X-ray/radio matched
population we use the C-statistic technique  (Cash et al. 1979) as
implemented in {\sc xspec} v11.2. This method has
been developed to extract information from X-ray spectra with small
number of counts. The data are grouped to have a minimum of one count
per bin but using higher binning factors does not modify the results. 
Assuming an  absorbed power law model we attempt to constrain the
$\rm N_H$ fixing  the power law  index to $\rm \Gamma=1.9$.  Only two
of the sources presented here (sources \#11 and \#12) show evidence
for enhanced  absorbing  columns $\rm >5\times10^{21}\,cm^{-2}$. The
remaining sources have X-ray spectral properties consistent with
Galactic absorption ($2\times10^{20}\rm cm^{-2}$). The results are
shown in column 5 of Table \ref{tbl3}.  We also attempt to constrain
the power law index  $\Gamma$ keeping the column density fixed to the
Galactic value. The results are also presented in Table
\ref{tbl3}. With the exception of sources \#11 and \#12 the present
sample has steep X-ray spectra.  Source \#5 in  Tables \ref{tbl2} and
\ref{tbl3} has sufficient counts that allow standard $\chi^{2}$
spectral analysis. This source is also fit by an absorbed power law
model with both the column density, $\rm N_H$, and the X-ray spectral
index, $\Gamma$, free parameters.  


Table \ref{tbl3} presents the X-ray properties of the
present sample. We list:    

{\bf 1.} source identification number.

{\bf 2.}  0.5-8\,keV net source counts.

{\bf 3.}  0.5-8\,keV X-ray flux in $\rm erg\,s^{-1}\,cm^{-2}$. 

{\bf 4.} 0.5-8\,keV X-ray luminosity in $\rm erg\,s^{-1}$, if a redshift
is available.

{\bf 5.} 1.4\,GHz radio luminosity density in $\rm W\,Hz^{-1}$, if a redshift
is available.

{\bf 6.} column density $\rm N_H$ estimated by the C-statistic for a
fixed power law index $\Gamma=1.9$. The only exception is source \#5
which is fit by an absorbed power law using $\chi^2$ statistics.

{\bf 7.} power law spectral index  $\rm \Gamma$ estimated by the
C-statistic for a fixed column density $\rm
N_H=2\times10^{20}\,cm^{-2}$. Source \#5 is fit using $\chi^2$
statistics.  

The sources in Tables \ref{tbl2} and \ref{tbl3} span both the radio
loud (\#1, 4, 5, 9, 10, 11, 12) and the radio quiet (\#2, 3, 6, 7, 8)  
regime with the radio-loudness parameter defined as in Stocke et
al. (1991; see Figure \ref{fig_aox} below). In  Table \ref{tbl3} both
radio quiet and loud systems have X-ray spectral index consistent on
average  with $\Gamma=1.9$. Coadding the X-ray spectra of radio loud 
and quiet sources separately yields $\Gamma=1.71^{+0.14}_{-0.14}$
and  $\Gamma=2.19^{+0.19}_{-0.33}$ respectively. Although radio loud
sources are flatter the spectral indices above are consistent within
the  90\%  confidence level. 

\begin{figure} 
\centerline{\psfig{figure=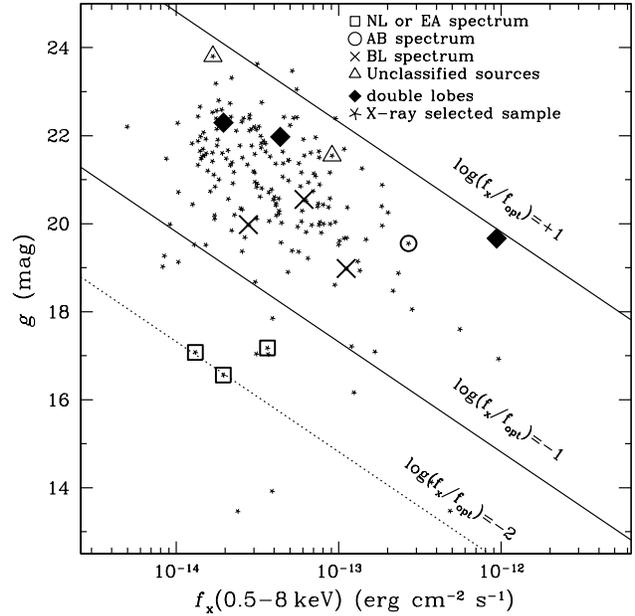,width=3.5in,angle=0}} 
\caption{
 $g$-band magnitude against 0.5-8\,keV flux. Small stars are
 0.5-8\,keV X-ray detections above the $5\sigma$ level. A large symbol
 on top of a star indicates a radio counterpart. Open squares are for
 sources with narrow emission line (NL) or narrow emission+absorption
 line (EA) optical spectra, open circles are
 for absorption line spectra (AB), crosses indicate broad line AGNs (BL)
 while triangles are for sources with no classification. Diamonds
 indicate radio sources with no spectroscopic information and double
 lobe morphology. One of the double lobe  sources is associated with
 an X-ray selected cluster. The lines indicate constant
 X-ray--to--optical flux ratios of +1, --1 and --2. The lines $\log
 f_X/f_{opt}=\pm1$ delineate the region of the parameter space
 occupied by powerful AGNs.  
 }\label{fig_fxfo}
\end{figure}

Figure \ref{fig_fxfo} plots $g$-band magnitude against 0.5-8\,keV
X-ray flux for both X-ray sources with radio counterparts and the
whole X-ray selected sample. The  $\log (f_X/f_{opt})=\pm1$ lines in
this figure delineate the region of the parameter space occupied by
AGNs. X-ray sources with radio emission span  a wide range of
X-ray--to--optical flux ratios. Although many X-ray/radio matches lie
in the AGN region of the parameter space, a small number of sources
have  $\log f_X/f_{opt}<-1$ suggesting  relatively optically luminous
obscured AGN and Low Luminosity AGNs (LLAGN; Lehmann et al. 2001) or
`normal' galaxies. 

The lines of constant X-ray--to--optical flux ratio in Figure
\ref{fig_fxfo} are estimated from the 0.5-8\,keV flux, 
$f_X(0.5-8\,{\rm keV})$, and the $g$-band magnitude according to
the relation
\begin{equation}\label{eq1}
\log\frac{f_X}{f_{opt}} = \log f_X(0.5-8\,{\rm keV}) +
0.4\,g +5.07.
\end{equation}
The equation above is derived from the X-ray--to--optical flux
ratio definition of Stocke et al. (1991) that involved 0.3-3.5\,keV
flux and $V$-band magnitude. These quantities are converted to
0.5-8\,keV flux and $g$-band magnitude assuming a mean colour
$B-V=0.8$ using the colour transformations of Fukugita et al. (1996)
and a power law X-ray spectral energy distribution with $\Gamma=1.8$.  

\begin{figure}
\centerline{\psfig{figure=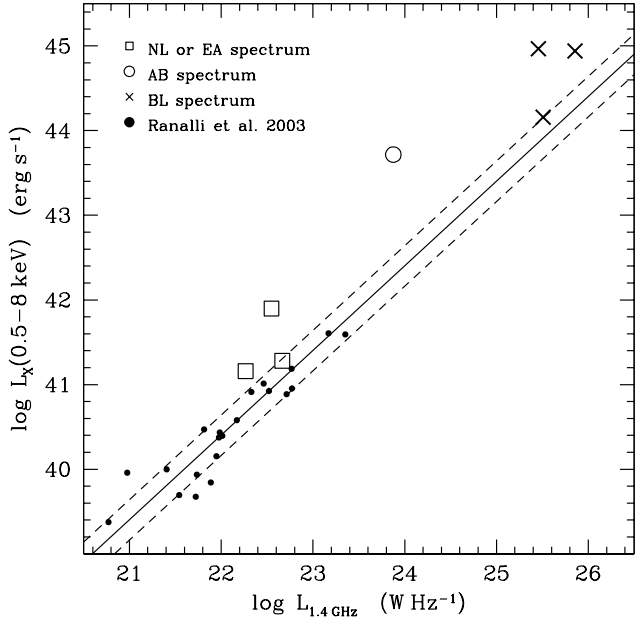,width=3.5in,angle=0}}
\caption
 {
 X-ray luminosity (0.5-8\,keV) against 1.4\,GHz radio luminosity
 density.  Open squares are for sources with narrow emission line or
 narrow emission+absorption optical spectra, open circles are for 
 absorption line spectra, stars indicate broad line AGNs. Filled
 circles are local star-forming galaxies from Ranalli et
 al. (2003). The continuous line is the best fit $L_X-L_{1.4}$
 relation for the sample of local star-forming galaxies of Ranalli et
 al. (2003). The dashed lines is the 1\,sigma rms scatter around the
 best fit. 	   
 }\label{fig_lxl14}
\end{figure}

Figure \ref{fig_lxl14} plots total band X-ray luminosity against
radio luminosity density for those sources in the present sample with
available redshift information. Both $L_X$ and $L_{1.4}$ are
k-corrected assuming a power law spectral energy distribution with
$\Gamma=1.8$ and $\alpha=0.8$ respectively.  Also shown is the best fit
$L_X-L_{1.4}$ relation for local star-forming galaxies derived by
Ranalli, Comastri \& Setti (2003). Two of our sample sources (\#7 and
\#8) have 
$L_X (\rm 0.5-8\,keV)\approx10^{41}\,erg\, s^{-1}\, cm^{-2}$ and lie
close to the Ranalli et al. (2003) relation suggesting they are
likely to be dominated by star-formation. These same sources also have
low X-ray--to--optical flux ratios in Figure \ref{fig_fxfo} 
($\log f_X/f_{opt}\approx-2$) which is additional evidence for
star-formation activity in these systems. The optical spectra for
these two sources are presented in Paper\,II. 
The remaining sources in Figure \ref{fig_lxl14} either deviate from
the $L_X-L_{1.4}$ relation of local star-forming systems (suggesting
the presence of AGN) or have X-ray luminosities well in excess of
typical starburst galaxies ($\approx10^{42}\rm\, erg\, s^{-1}$; Moran,
Lehnert \& Helfand 1999).  

We also find a double lobe radio source (\#1) associated with
diffuse X-ray cluster emission. As discussed in Appendix \ref{app1}
this cluster is likely to lie at $z\approx0.6$. Powerful radio
galaxies  are known to be good tracers of dense environments with
Fanaroff-Rilley I type sources (FRI)  associated with rich groups or
clusters and Fanaroff-Rilley II type sources (FRII) avoiding rich
clusters at low-$z$ (e.g. Zirbel 1997 and references
therein). Recently, Zanichelli et al. (2001) used NVSS data  to
compile an intermediate redshift ($z\approx0.1 - 0.3$) cluster sample
based on radio selection. Follow-up optical photometry  and
spectroscopy confirmed that a significant fraction of their targeted 
cluster candidates are real, demonstrating the strength of radio
selection in cluster studies. Our X-ray/radio matched sample comprises
a total of 3 double lobe radio sources (\#1, \#5 and \#12) of which
only one (\#1) is clearly associated with X-ray cluster emission. The
remaining two lie  in either poor or high-$z$ clusters/groups that
remain undetected in both the X-ray and the optical data. Indeed, as
discussed in Appendix \ref{app1} we do not find evidence for optical
galaxy overdensity in the vicinity of sources  \#5 and \#12, at least
to the optical magnitude limit of the SDSS. 

Sources with absorption line optical spectra and high X-ray
luminosities ($L_X\approx10^{42}-10^{43}\rm\, erg\,s^{-1}$), similar
to those found in the present sample (\#3, 4 in Table 1) , have also
been identified in previous X-ray surveys (e.g. Griffiths et al. 1995;
Blair, Georgantopoulos \& Stewart 1997; Allen, Di Matteo \&  
Fabian 2000; Comastri et al. 2002; Brusa et al. 2003). Possible
scenarios for the  
observed X-ray emission in these optically normal (e.g. no emission
lines) systems include (i) BL-Lac type activity (e.g. Blair et
al. 1997), (ii)  Advection Dominated Accretion Flows (ADAF; Narayan \&
Yi 1995; Di Matteo et al. 2000), (iii) diffuse cluster of group  X-ray
emission and (iv) LLAGN with its optical signature diluted by the
host galaxy light  (e.g. Moran, Filippenko \&  Chornock 2002;
Severgnini et al. 2003).  

For sources \#3, 4 we can exclude the possibility of X-ray cluster
emission. The X-ray sources associated with these two galaxies are not
extended. Also, no optical galaxy overdensity is found in the vicinity
of the two sources (Basilakos et al. 2004, Paper\,III; Goto et
al. 2002). The  
evidence above is against the cluster emission scenario although the
presence of hot gas associated with a group (that remains undetected
in both the X-ray and the optical data) cannot be ruled out. 
Below we argue that source \#3 is likely to be associated with an ADAF
or a LLAGN, while  source \#4 is a BL-Lac candidate.

We explore the nature of these sources using their
two-point spectral indices, optical--to--X-ray  ($\alpha_{OX}$) and 
radio--to--optical  ($\alpha_{RO}$) defined by Stocke et  al. (1991): 
\begin{equation}\label{eq3}
 \alpha_{OX}= - \log (f_{\rm 2\,keV}/f_{\rm 2500})/2.605,
\end{equation}
\begin{equation}\label{eq4}
 \alpha_{RO}= \log (S_{\rm 5\,GHz}/f_{\rm 2500})/5.38,
\end{equation}
where $f_{\rm 2\,keV}$, $S_{\rm 5\,GHz}$ and $f_{\rm 2500}$ are the
monochromatic fluxes at 2\,keV, 5\,GHz and $\rm 2500\AA$
respectively. The  $S_{\rm 5\,GHz}$  is estimated from the  1.4\,GHz
flux density assuming a power law spectral energy distribution of the
form $f_{\nu}\propto \nu^{-\alpha}$ with $\alpha=0.8$.
The X-ray flux density at 2\,keV is estimated from
the 0.5-8\,keV flux  assuming a mean photon spectral index
$\Gamma=1.8$. The monochromatic optical flux at 2500\,\AA,  $f_{\rm
2500}$, is estimated from the SDSS $u$-band magnitudes assuming an
optical power law spectral slope $\alpha=0.5$ (Brinkmann et al. 2000).
The flux densities are not k-corrected but this effect is
small and does not modify the estimated monochromatic flux ratios. 

The results are shown in Figure \ref{fig_aox} plotting $\alpha_{OX}$
against $\alpha_{RO}$. Also shown are the regions of the parameter
space occupied by different classes of objects as defined by Stocke et
al. (1991; see their Figure 6b). The absorption line  galaxy in our
sample (\#4) has $\alpha_{OX}$ and $\alpha_{RO}$ consistent with
those of BL-Lacs. This coupled with the soft X-ray spectral properties
of this system and the absorption line optical spectrum strongly
support the BL-Lac scenario. Fossati et al. (1998) showed that the
peak frequency of the BL-Lac SED is a function of the radio luminosity
of these objects: less luminous systems peak at UV and soft X-ray
wavelengths (High energy peak BL-Lacs), while more luminous radio
sources have a peak at the infrared regime (Low energy peak
BL-Lacs). Source \#4 at $z=0.240$ has a radio luminosity at 5\,GHz of 
$\log \nu L_\nu\approx10^{40}\,\rm erg\,s^{-1}$ suggesting that this
is a  High energy peak BL-Lac (see Figure 7a of Fossati et
al. 1998). The estimated radio luminosity is also consistent with the
steep X-ray spectral index of this source in Table 2 (see Figure 11
of Fossati et al. 1998). We also note that absorption line systems
with properties similar to those of source \#4 have  been identified
in previous X-ray  surveys with available radio observations (Gunn et
al. 2003; Brusa et al. 2003).   

Three sources in Figure \ref{fig_aox} lie close to the `normal' galaxy
region of the Stocke et al. (1991) classification scheme. Two of these
sources are the `normal' galaxy candidates showing narrow emission
line optical spectra. The third source is \#3 in Tables 1 and 2
exhibiting  both narrow emission and absorption optical lines. The
evidence above combined with the fact that this system deviates from
the Ranalli et al. (2003) $L_X-L_{1.4}$ relation in Figure
\ref{fig_lxl14} suggest an ADAF or a LLAGN. 

Broad optical emission line systems and double lobe sources  in Figure
\ref{fig_aox} occupy the radio loud and the radio quiet AGN space. We
note that the one double lobe  source associated with the  X-ray
cluster is not plotted here since its X-ray emission is not due to the
central AGN.  One of the two spectroscopically unclassified sources
(\#11) in the present sample has  $\alpha_{OX}$ and $\alpha_{RO}$
consistent with those of BL-Lacs. This is inconsistent however, with
the relatively hard X-ray spectral properties of this source in Table
2 suggesting $\rm N_H\approx1.5\times10^{22}$ for
$\Gamma=1.9$. The second  spectroscopically unclassified source (\#10)
lies in the border line between radio loud QSOs and BL-Lacs.

\begin{figure}
\centerline{\psfig{figure=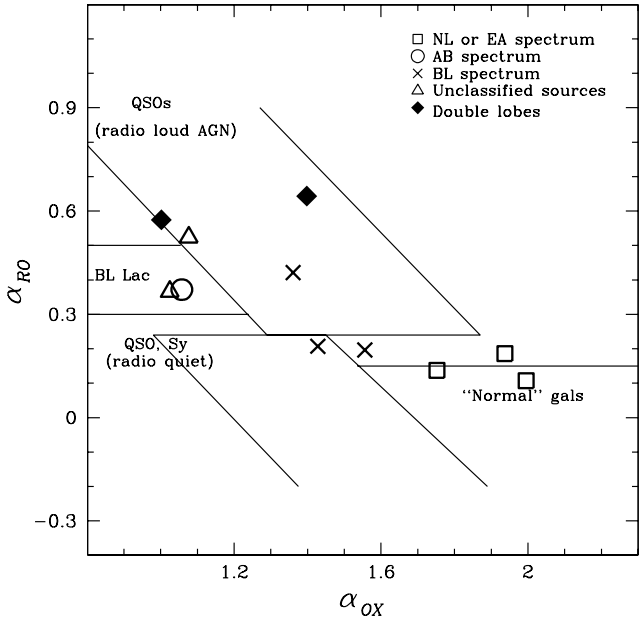,width=3.5in,angle=0}}
\caption
 {$\alpha_{RO}$ against $\alpha_{OX}$ for the XMM-{\it Newton}/2dF
 survey sources with radio counterparts in the FIRST survey. The
 symbols are the same as in Figure \ref{fig_fxfo}. The one double lobe
 source associated with the  X-ray cluster is not plotted here since
 its X-ray emission is not due to the central AGN.   
 }\label{fig_aox}
\end{figure}

A detailed description of individual sources is presented in
Appendix 
\ref{app1}. The classification of the present sample into different
classes is performed  on the basis of their optical spectroscopic
(i.e. spectral features), photometric (i.e. resolved or point-like
sources) and X-ray properties (i.e. X-ray luminosity,
X-ray--to--optical flux ratio, X-ray spectral properties). More
information about the 
source classification can be found in  Appendix
\ref{app1}. The present sample of X-ray/radio matches comprises: 
{\bf  (i)} 3 broad emission line AGNs (\#2, 6, 9), 
{\bf (ii)} 2 sources with double  lobe radio morphology also
indicating AGN activity (\#5, 12), {\bf (iii)} 1 double lobe source (\#1)  
associated with X-ray cluster emission, {\bf (iv)} 2 systems with
optical spectra dominated by the host galaxy (i.e. absorption and/or
narrow emission lines) and X-ray/optical properties  suggesting AGN
activity (\#3, 4),  {\bf (v)} 2 `normal' galaxy candidates exhibiting
narrow emission line optical spectra (\#7, 8) and  
{\bf  (vi)} 2 unclassified sources with no optical spectroscopic
information (\#10, 11). The  X-ray/optical properties of this latter
class of sources suggests AGN activity. We also note that although the
X-ray and optical properties of `normal' galaxy candidates are
consistent with stellar origin for the X-ray emission we cannot
exclude the possibility of heavily obscured AGN or LLAGN.  
Summarising the classification above, 9 out of the 12 X-ray/radio
matches are associated  with AGN activity on the basis of the broad
optical  emission lines (total of 3),  radio morphology (2) or 
X-ray/optical properties (4).  One radio source is associated with
X-ray cluster emission with the remaining two sources being `normal'
galaxy candidates.

\section{Discussion}\label{discussion}
Brinkmann et al. (2000) cross-correlated the FIRST radio survey with
the RASS-II catalogue with a 0.1--2.4\,keV limiting flux of 
$\approx10^{-13}\,\rm erg\,s^{-1}\,cm^{-2}$. These authors found that 
$\approx30\%$ of the RASS-II sources have radio  counterparts to the
limit of the FIRST survey. In the present survey probing X-ray fluxes
about 1\,dex fainter than the RASS-II we find that 12 out of 291 X-ray
sources ($\approx4\%$) have radio counterparts. This fraction is much
lower than that of the RASS-II ($\approx30\%$) suggesting that a
significant fraction of the X-ray sources have radio emission below
the FIRST flux density limit. Ciliegi et al. (2003) suggest that the
fraction of the X-ray/radio matches strongly depends on the relative
depths of the X-ray and the radio data. They quantify this effect using
the ratio of the X-ray--to--radio flux density limits of different
surveys, $S_{\rm 1.4\,GHz}/f_{\rm 2 keV}$. For the XMM-{\it
Newton}/2dF survey we estimate $S_{\rm 1.4\,GHz}/f_{\rm 2
keV}\approx10^{6}$ (assuming $\Gamma=1.8$) which is 1\,dex higher than
that of $\approx10^{5}$ for the Brinkmann et al. (2000)
sample. Ciliegi et al. (2003) also provide a comparison of the
X-ray/radio identification rate of various surveys  spanning a range
of X-ray and radio limits (see their Table 2). The XMM-{\it
Newton}/2dF survey  has the lowest fraction of X-ray/radio matched
sources in this table (4\%) and the highest  $S_{\rm 1.4\,GHz}/f_{\rm
2 keV}$ ratio, while the FIRST/RASS-II sample of Brinkmann et
al. (2000) is comparable with other soft X-ray selected surveys with
similar $S_{\rm 1.4\,GHz}/f_{\rm 2 keV}$ ratios (e.g. Stocke et
al. 1991).    

In addition to soft X-ray samples a number of studies have explored
the radio properties of hard  sources finding much higher X-ray/radio
identification rates than those in softer  X-ray surveys (Barger et
al. 2001; Ciliegi et al. 2003). This is
attributed to both observational effects  (e.g. deeper radio data) and
the insensitivity of both the radio and the hard X-ray wavelengths  to
obscured AGNs  (e.g. Ciliegi et al. 2003). Moreover,  hard X-ray
selected AGNs with radio counterparts have harder X-ray spectra
(i.e. more absorbed) than non-radio detected AGNs (e.g. Bauer  
et al. 2002; Georgakakis et al. 2004b). It has been suggested that
this may be due to the presence of circumnuclear starbursts in radio
detected AGNs that both feed and obscure the central black hole
(e.g. Bauer et al. 2002). Only two sources in the present sample
(\#11, 12) show evidence for absorption with the remaining having soft
X-ray spectral properties consistent with little or no photoelectric
obscuration. These two sources are also detected in the hard X-ray
spectral band (2-8\,keV) above the $6\sigma$ detection threshold
(Georgantopoulos et al. 2004, Paper\,IV).   

Brinkmann et al. (2000) also present follow up spectroscopic
observations for about half of their X-ray/radio matched
population. They find that $\approx71\pm5\%$ (324/454) of  the sources
are AGNs, $\approx22\pm2\%$ (99/454) are early type galaxies with
absorption line spectra, $\approx5\pm1\%$ (23/454) are classified
starbursts and $\approx2\pm1\%$ (8/454) are clusters. We note that the
class of  absorption line galaxies in the Brinkmann et al. (2000)
sample is likely to be  contaminated by AGNs that do not reveal their
presence in the optical spectrum (Severgnini et al. 2003).   

In our sample about  $75\pm33\%$ (9/12) of the X-ray/radio matches are
AGNs on the basis of their optical spectra, radio morphology or
X-ray/optical properties. Starforming `normal' galaxy candidates
represent $\approx17\pm13\%$ (2/12) of the X-ray/radio poulation while
we find 1  radio source associated with  cluster X-ray
emission. We find a larger fraction of starforming galaxy
candidates than Brinkmann et al. (2000) that can be interpreted as
tenuous evidence for increasing fraction of `normal' galaxies with
decreasing X-ray flux. One should be cautious however, since  the
estimated fractions above  are clearly affected by small number
statistics that do not allow firm conclusions to be
drawn.  

`Normal' galaxy candidates have only recently been reliably identified
in X-ray surveys suggesting that their number density increases with
decreasing X-ray flux (Bauers et al. 2002; Hornschemeier et al. 2003;
Kim et al. 2004; Paper\,II).  On the contrary, {\it
radio} selection below 10\,mJy has long been known to be an efficient
tool for identifying starburst galaxies albeit with some AGN
contamination (Georgakakis et al. 1999; Magliocchetti et al. 2002;
Chapman et al. 2003). Starforming systems appear in increasing numbers
with decreasing flux density below 10\,mJy and are believed to
dominate the sub-mJy radio source counts (Hopkins et al. 1998;
2003). The two `normal' galaxy candidates found in the present study
by cross-correlating the X-ray and radio samples have also been
identified in Paper\,II on the basis of both their
low   $f_X/f_{opt}$ and their optical spectral properties. Therefore,
radio selection to the limit of the FIRST survey has identified {\it
all} the X-ray {\it selected} `normal' galaxy candidates in our
XMM-{\it Newton}/2dF survey. Although the statistics are poor the
evidence above suggests that radio selection can potentially single
out `normal' starforming galaxies within X-ray samples.    

Magliocchetti et al. (2002) studied the nature of the optically bright
($b_j<19.45$\,mag) FIRST radio sources using optical spectra from the
2dFGRS. They find that as much as 32\% of their spectroscopic sample
are star-forming systems with the rest being early type galaxies or
Seyfert 1 and 2 type systems. The majority of star-forming
galaxies are found at relatively low redshifts, $z<0.1$. 

From the 72 FIRST radio sources overlapping with the XMM-{\it
Newton}/2dF survey (see section 2.2), a total of 18 have optical
counterparts on the SDSS brighter than $g\approx19.2$\,mag
(i.e. similar to $b_j=19.45$\,mag). On the basis of the Magliocchetti
et al. (2002)  study from the above 18 radio sources brighter than
$g\approx19.2$\,mag we expect about 6 star-forming galaxies. This
number can be compared with the two {\it X-ray}  selected ``normal''
galaxy candidates identified both in the present study and in
Paper\,II. Although the statistics are poor the analysis above
suggests that  to the limit of the XMM-{\it Newton}/2dF survey we have
only identified a sub-sample of the total population of {\it radio}
selected starforming galaxies (to the limit  $g=19.2$\,mag).  

\begin{figure}
\centerline{\psfig{figure=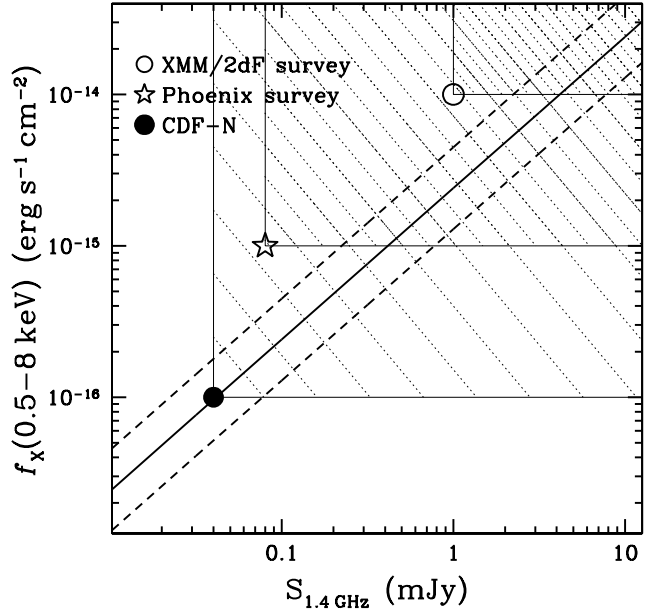,width=3.5in,angle=0}}
\caption
 {
 1.4\,GHz radio flux density against X-ray flux in the 0.5-8\,keV
 spectral band. The continuous line is the  $f_{\rm X}-S_{\rm 1.4\,GHz}$
 relation of local starforming galaxies derived by Ranalli et
 al. (2003; see text for details). The dashed lines represent the 
 $1\sigma$ rms uncertainty of the $L_{\rm X}-L_{\rm 1.4\,GHz}$
 relation. The symbols represent the X-ray flux and radio flux density
 limits of various X-ray/radio surveys: the open circle is the XMM-{\it Newton}/2dF
 survey with limiting fluxes $f_X(\rm 0.5-8\,keV)=10^{-14}\rm
 erg\,s^{-1}\,cm^{-2}$ and $S_{\rm 1.4\,GHz}=1\rm\, mJy$, the star is
 for the Phoenix/XMM survey (Georgakakis et al. 2003b; $f_X(\rm
 0.5-8\,keV)=10^{-15}\rm  erg\,s^{-1}\,cm^{-2}$, $S_{\rm
 1.4\,GHz}=0.08\rm\, mJy$) and the filled circle represents the CDF-N
 (Bauers et al. 2002; $f_X(\rm 0.5-8\,keV)=10^{-16}\rm
 erg\,s^{-1}\,cm^{-2}$, $S_{\rm 1.4\,GHz}=0.05\rm\, mJy$). The region
 of the parameter space (X-ray flux and radio flux density) accessible
 by  these datasets is marked by the dotted diagonal lines. 
 }\label{fig_fxs14}
\end{figure}

It is an interesting exercise to quantify the flux limits
that radio and X-ray surveys should have to detect the same
population of starforming galaxies. We use the $L_X-L_{1.4}$
relation of local star-forming galaxies derived by Ranalli et
al. (2003) to establish the relation between  X-ray flux, $f_X(\rm
0.5-8\,keV)$, and radio flux density, $S_{\rm 1.4\,GHz}$. Since
the  Ranalli et al. (2003)  $L_X-L_{1.4}$ relation is linear one can
factor the distance out to obtain an $f_X-S_{1.4}$ relation for
starforming galaxies. This simple transformation however, ignores the 
k-correction. The effect of the  differential X-ray/radio
k-correction is  nevertheless small compared to the $1\sigma$ rms
scatter of the Ranalli et al. (2003) $L_X-L_{1.4}$ relation. The
derived $f_X(\rm 0.5-8\,keV)-S_{1.4\,\rm GHz}$  relation for
star-forming galaxies is compared in Figure \ref{fig_fxs14} with the
limits of various X-ray/radio surveys including the XMM-{\it
Newton}/2dF survey. The region of the parameter space
(e.g. X-ray/radio flux range) accessible by these surveys is also 
marked in this figure.  We conclude that our X-ray survey is too  
shallow to detect the radio selected star-forming population to the
limit of the FIRST survey. An order of magnitude improvement in
sensitivity at X-ray wavelengths is essential for such a
study. Nevertheless, Figure \ref{fig_fxs14} suggests that combining
X-ray and radio surveys with carefully selected limiting fluxes is a
powerful tool for identifying star-forming galaxies. In particular,
this figure shows that the CDF-N has X-ray/radio flux limits that
allow a large fraction of the   radio/X-ray selected star-forming galaxy
population to be identified at both wavelengths. Indeed,  Bauer et
al. (2002) investigated the association between faint X-ray and radio
source populations detected in the 1\,Ms CDF-N. They found that the
majority of X-ray sources with narrow emission-line spectra also have
$\mu$Jy radio counterparts and are likely to be dominated by
star-formation activity. 

\section{Conclusions}\label{conclusions}

In this paper we explore the nature of the X-ray sources with radio
counterparts by cross-correlating the FIRST 1.4\,GHz radio survey with
a wide field ($\rm 1.6\,deg^{2}$) shallow
[$f_X\rm (0.5 - 8\,keV)\approx 10^{-14} \, erg \, s^{-1}$] XMM-{\it
Newton} survey. Our sample comprises 12 X-ray/radio matches
representing about 4\% of the X-ray selected sample. The
majority of these sources (9/12) have properties suggesting 
AGN activity. Two sources in the sample have narrow emission line
optical spectra, X-ray luminosities $L_X\rm \approx
10^{41}\,erg\,s^{-1}$, radio powers $L_{1.4}\rm \approx
5\times10^{22}\,W\,Hz^{-1}$ and X-ray--to--optical flux ratios
$\approx-2$ suggesting `normal' galaxies powered by star-formation
activity. Finally we find 1 double radio source associated with
X-ray cluster emission. Both radio loud and quiet systems in the
present sample  have mean X-ray spectral properties consistent with 
$\Gamma\approx1.9$.     

We also argue that radio selection to the limit of the FIRST survey has
identified {\it all} the X-ray {\it selected} `normal' galaxy 
candidates in our XMM-{\it Newton}/2dF survey. Although small number
statistics hamper a secure interpretation the evidence above
suggests that radio selection can potentially single out `normal'
starforming galaxies within X-ray samples. Using the $L_X-L_{\rm
1.4\,GHz}$ relation for local starforming galaxies we quantify the
flux limits that X-ray and radio surveys should have to identify the
same population of starbursts.

\section{Acknowledgments}
 We thank the anonymous referee for useful comments and suggestions
 that improved this paper. This work is jointly funded by the European
 Union and the Greek Government  in the framework of the programme
 ``Promotion of Excellence in Technological Development and Research'',
 project ``X-ray Astrophysics with ESA's mission XMM''.  The XMM-{\it
 Newton}/2dF  survey data as well as part of the observations
 presented here are electronically available at {\sf
 http://www.astro.noa.gr/$\sim$xray/}. 

 We acknowledge use of the 100k data release of the 2dF Galaxy   
 Redshift Survey. The 2dF QSO Redshift Survey (2QZ) was compiled by
 the 2QZ survey team from observations made with the 2-degree Field on
 the Anglo-Australian Telescope.  

 Funding for the creation and distribution of the SDSS Archive has
 been provided by the Alfred P. Sloan Foundation, the Participating
 Institutions, the National Aeronautics and Space Administration, the
 National Science Foundation, the U.S. Department of Energy, the
 Japanese Monbukagakusho, and the Max Planck Society. The SDSS Web
 site is http://www.sdss.org/. The SDSS is managed by the
 Astrophysical Research Consortium (ARC) for the Participating
 Institutions. The Participating Institutions are The University of
 Chicago, Fermilab, the Institute for Advanced Study, the Japan
 Participation Group, The Johns Hopkins University, Los Alamos
 National Laboratory, the Max-Planck-Institute for Astronomy (MPIA),
 the Max-Planck-Institute for Astrophysics (MPA), New Mexico State
 University, University of Pittsburgh, Princeton University, the
 United States Naval Observatory, and the University of Washington. 
 
\appendix 
\section{Notes on individual sources}\label{app1}

{\bf FIRST\,J134304.6--000055:} this double lobe radio source has
optical counterpart lying $\approx1$\,arcsec from the radio
centroid. Although optical spectroscopy is not available for this
source the radio morphology indicates AGN activity. The  
X-ray source likely to be associated with FIRST\,J134304.6--000055
lies 3\,arcsec from the radio centroid. The X-ray emission is
extended suggesting hot gas cluster emission. Indeed, Couch et
al. (1991) used photographic material to identify an optical galaxy
overdensity in the vicinity of the X-ray source. Follow-up
spectroscopy of selected galaxies in this field by these authors
suggests a cluster redshift of $\approx0.6$. In Paper\,III we 
use CCD photometry  to identify a statistically significant optical
galaxy overdensity in the vicinity of the X-ray source. Using
photometric methods (e.g. Postman et al. 1996) these authors estimate
a redshift $z\approx0.6$ for this cluster candidate.

{\bf FIRST\,J134233.0--001553:} the radio centroid lies
$\approx3$\,arcsec off the X-ray position and $\approx0.5$\,arcsec
from the optical counterpart. Optical spectroscopy shows that this
source is associated with a broad emission line QSO at a redshift
$z=2.132$.   

{\bf FIRST\,J134212.2--001737:} the radio position coincides with
that of a  $g\approx17.0$\,mag galaxy at $z=0.087$. The X-ray centroid is
offset from both the optical and radio source positions by
4\,arcsec. This may suggest that the X-ray source may not be
associated with either the optical galaxy or the radio source. Keeping
this caveat in mind in what follows we assume that both the radio and
the X-ray  emission originate from the optical galaxy at $z=0.087$.   
The 2dFGRS optical spectrum of this systems suggest an early type
spiral with both emission ($\rm H\alpha$, $\rm [N\,II]\,6583\AA$, $\rm
[O\,II]\,3727\AA$) and absorption (H+K, $\rm H\beta$, $\rm H\gamma$,
$\rm H\delta$, $\rm NaD\,5893\AA$) lines. The X-ray--to--optical flux
ratio of $-1.5$ and the X-ray luminosity,
$L_X(0.5-8)\approx8\times10^{41}\rm\,erg\,s^{-1}$,  suggest either AGN 
or X-ray properties dominated by stellar processes. In 
Figure \ref{fig_lxl14} this systems deviates from the mean
$L_X-L_{1.4}$ of local star-forming galaxies favoring the AGN
scenario for the observed X-ray emission. This source also lies close
to the `normal' galaxy region of the $\alpha_{OX}-\alpha_{RO}$ diagram 
in Figure \ref{fig_aox}. The C-statistic X-ray spectral analysis
suggests soft X-ray spectral properties consistent with Galactic
column density absorption. This source is also discussed in Paper\,II
(their source TGN\,336Z232). We classify this source LLAGN or ADAF. 

{\bf FIRST\,J134347.5+002024:} the X-ray and radio positions
coincide within 3\,arcsec. The optical counterpart is likely to be 
a $g\approx19.5$\,mag galaxy at a redshift $z=0.240$. The optical
spectrum of this systems does not show emission lines suggesting
an early type galaxy. The X-ray--to--optical flux ratio of $+0.3$ and
the X-ray luminosity,
$L_X(0.5-8)\approx5\times10^{43}\rm\,erg\,s^{-1}$,  suggest AGN
activity. The two point spectral indices of this source, $\alpha_{OX}$
and $\alpha_{RO}$,  are consistent with those of BL-Lacs. This
coupled with  the soft X-ray spectral properties of this system and
the absorption line optical spectrum strongly support the BL-Lac
scenario.  

{\bf FIRST\,J134414.2+001642:} this radio source has double lobe
radio morphology indicating AGN activity.  The 
X-ray source likely to be associated with  FIRST\,J134414.2+001642
lies 1\,arcsec from the radio centroid. The optical counterpart has  
$g\approx19.7$\,mag and no optical spectroscopic information.  The
X-ray--to--optical flux ratio of $+0.9$ is consistent with AGN
activity.  Our X-ray spectral analysis suggests soft X-ray spectrum.

{\bf FIRST\,J134232.4--003151:} the X-ray and radio positions
overlap within less than 1\,arcsec. The radio source is associated
with a  $g=19.6$\,mag QSO at $z=1.209$. The X-ray--to--optical flux 
ratio of $-0.3$ and the X-ray luminosity
$L_X(0.5-8)\approx8\times10^{44}\rm\,erg\,s^{-1}$ are also consistent 
with AGN activity. This is the only source that has enough photons
allowing detailed X-ray spectral analysis. The X-ray spectrum is
best fit by a power law spectrum with  $\Gamma\approx1.8$ and
photoelectric extinction $\rm N_H<1.4\times10^{20}\,cm^{-2}$
consistent with the  Galactic value.

{\bf FIRST\,J134133.4--002432:} the radio source lies 3\,arcsec off
the X-ray position. The most probable optical counterpart of both the
X-ray and radio sources is a $g\approx17.1$\,mag galaxy. The X-ray
centroid is offset by 3\,arcsec from the 
optical centre of the galaxy. This is a narrow  emission line system
at $z=0.0717$ showing $\rm [O\,III]\,4959+5007\AA$ doublet, $\rm
H\alpha$ and $\rm [N\,II]\,6583\,\AA$. Although the signal--to--noise
ratio of the spectrum is low, $\rm H\beta$ is not visible in
emission. This coupled with the strong $\rm [O\,III]\,5007\AA$ feature
may suggest a Seyfert 2 type system. 
Moreover, the low  X-ray luminosity 
$L_X(0.5-8)=1.9\times10^{41}\rm\,erg\,s^{-1}$ and
X-ray--to--optical flux ratio $\approx-2.0$ are consistent with a
`normal' galaxy powered by stellar processes. In
Figure \ref{fig_lxl14} this system lies close to the mean
$L_X-L_{1.4}$ of local star-forming galaxies favoring the
star-formation as a probable source of the observed X-ray emission.

{\bf FIRST\,J134137.7--002555:} The X-ray and radio positions
coincide within 3\,arcsec. The most probable optical counterpart of
both the X-ray and radio sources is a 
$g\approx16.6$\,mag galaxy at $z=0.052$. This narrow emission line
galaxy is classified as H\,II on the basis of its optical spectral
properties (Terlevich et al. 1991). Its X-ray luminosity 
$L_X(0.5-8)=1.5\times10^{41}\rm\,erg\,s^{-1}$ and X-ray--to--optical
flux ratio $\approx-2.0$ also suggest `normal' galaxy. Our X-ray
spectral analysis also suggests soft X-ray properties. 
In Figure \ref{fig_lxl14} this system lies close to
the mean $L_X-L_{1.4}$ of local star-forming galaxies also favoring
the star-formation scenario for the X-ray emission.

{\bf FIRST\,J134412.9--003006:} the X-ray and radio positions
coincide within 3\,arcsec. The most probable optical counterpart for
both the X-ray and the radio source is a $g\approx20.6$\,mag broad
line AGN at $z=0.708$.  Its X-ray luminosity
$L_X(0.5-8)=1.4\times10^{44}\rm\,erg\,s^{-1}$ and X-ray--to--optical
flux ratio $\approx+0.1$ also consistent with AGN activity. The
C-statistic suggests $\rm N_H$ consistent with the Galactic column
density.  

{\bf  FIRST\,J134431.8--002832:} the radio source lies 3\,arcsec off
the X-ray position. The most probable optical counterpart is a
$g\approx23.8$\,mag galaxy. No optical spectroscopic information is
available for this source. The X-ray--to--optical flux ratio of $+0.82$
suggests AGN activity. The X-ray spectrum is very steep
$\Gamma\approx2.3$ for $\rm N_H=2\times10^{20}\,cm^{-2}$. In the
$\alpha_{OX} - \alpha_{RO}$ diagram this source
lies in the borderline between radio loud AGNs and BL-Lacs.    

{\bf  FIRST\,J134128.4--003120:} the radio source lies 3\,arcsec off
the X-ray centroid. The most probable optical counterpart is a
$g=21.55$\,mag galaxy. Although no optical spectroscopic information
is available for this system the X-ray--to--optical flux ratio of
$+0.65$ suggest AGN activity. This source lies in the BL-Lac region of
the  $\alpha_{OX} - \alpha_{RO}$ parameter space. The C-statistic
suggests hard X-ray spectral properties with photoelectric absorption
$\approx1.5\times  10^{22}\,\rm cm^{-2}$ ($\Gamma=1.9$) which is
inconistent with a BL-Lac. 

{\bf FIRST\,J134447.0--003009:} The X-ray and radio positions
coincide within less than $1$\,arcsec. The radio source has double lobes
indicating the presence of an AGN.  The optical
counterpart has $g\approx22.3$\,mag and no optical spectral
information are available. The X-ray--to--optical flux
ratio of $+0.65$ also suggest AGN activity. The C-statistic suggests
hard X-ray spectral properties with photoelectric absorption
$\ga10^{22}\,\rm cm^{-2}$ ($\Gamma=1.9$).

\end{document}